\documentclass[fleqn,10pt]{SelfArx} 

\usepackage{lipsum} 

\definecolor{color1}{RGB}{0,0,0} 
\definecolor{color2}{RGB}{0,0,0} 

\usepackage{hyperref} 
\hypersetup{hidelinks,colorlinks,breaklinks=true,urlcolor=color2,citecolor=color1,linkcolor=color1,bookmarksopen=false,pdftitle={Title},pdfauthor={Author}}

\Archive{} 

\PaperTitle{Simple and reliable method of conductive SPM probe fabrication using carbon nanotubes} 

\Authors{Vyacheslav Dremov\textsuperscript{1,2}*, Vitaly Fedoseev\textsuperscript{1}, Pavel Fedorov\textsuperscript{1,2}, Artem Grebenko\textsuperscript{1,2}} 
\affiliation{\textsuperscript{1}\textit{Institute of Solid State Physics, RAS, 142432 Chernogolovka, Russian Federation}} 
\affiliation{\textsuperscript{2}\textit{Interdisciplinary Center for Basic Research, Moscow Institute of Physics and Technology, 141700 Dolgoprudniy, Russian Federation.}} 
\affiliation{*\textbf{Corresponding author}: dremov@issp.ac.ru} 

\newcommand{\keywordname}{Keywords} 

\Abstract{ \em We demonstrate the procedure of Scanning Probe Microscopy (SPM) conductive probe fabrication with a single multi-walled carbon nanotube (MWNT) on a silicon cantilever pyramid. The nanotube bundle reliably attached to the metal-covered pyramid is formed using electrophoresis technique from the MWNT suspension. It's shown that the dimpled aluminium sample can be used both for shortening/modification of the nanotube bundle by applying pulse voltage between the probe and the sample, and for controlling the probe shape via Atomic Force Microscopy (AFM) imaging the sample. It allows to fabricate a probe suitable for SPM imaging in the contact and modulation regimes. The majority of such probes are conductive with conductivity not degrading within hours of SPM imaging.}

\begin{document}

\flushbottom 

\maketitle 

\tableofcontents 

\thispagestyle{empty} 


\section{Introduction} 
\begin{figure}[tbp]
       \centering
        \begin{subfigure}[]{\linewidth}
               \centering
                \includegraphics[width=0.6\linewidth]{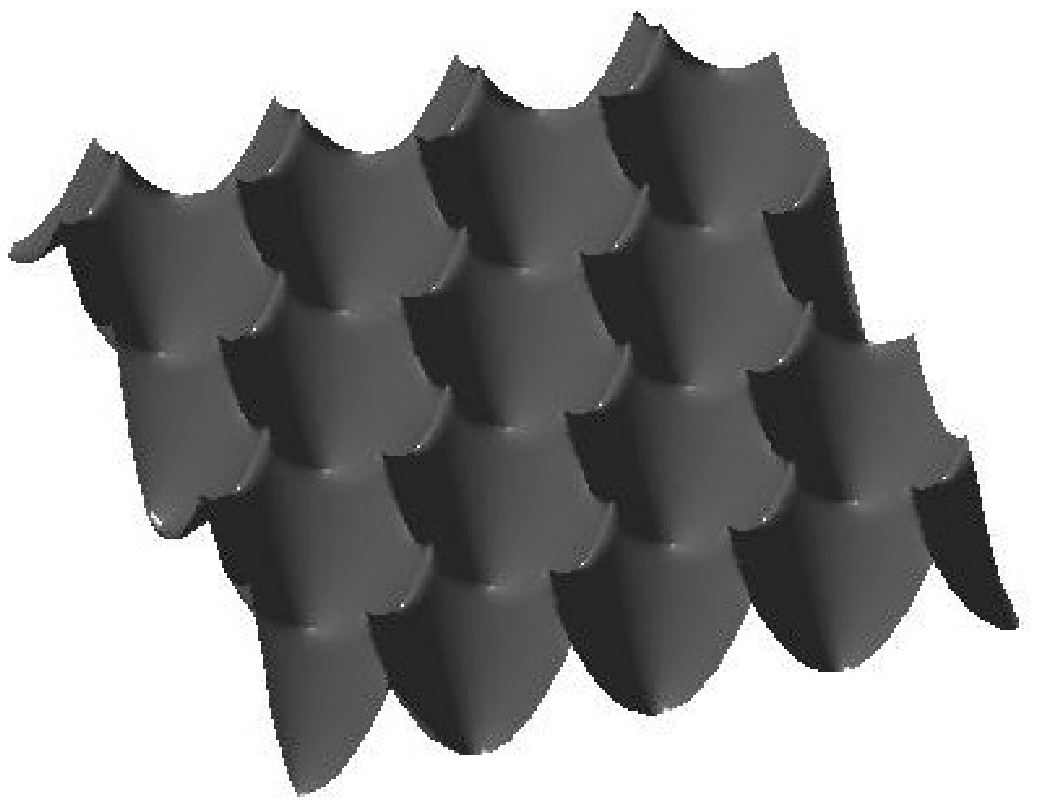}
                \caption{}
                \label{fig:fig1a}
        \end{subfigure}%
        
        \begin{subfigure}[]{\linewidth}
        \centering
                \includegraphics[width=0.6\linewidth]{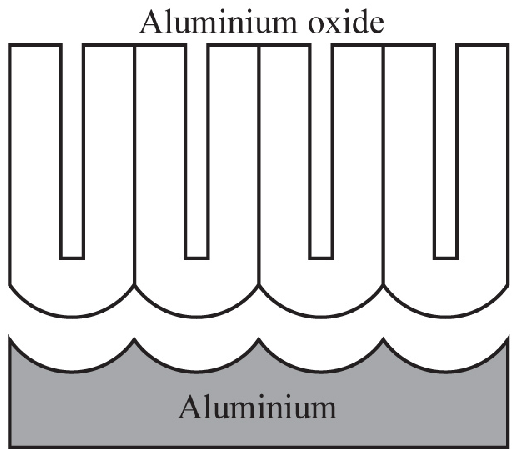}
                \caption{}
                \label{fig:fig1b}
        \end{subfigure}
        \caption{(a)Model of the dimpled aluminium. (b) The dimpled aluminium is produced by selective removal of alumina obtained by anodization of aluminium.}
        \label{fig:fig1}
\end{figure}

The problem of the appropriate probe fabrication for nanostructures investigation emerged together with the origin of Scanning Probe Microscopy. One can consider a rigid high aspect ratio probe having a single atom at its tip to be an ideal one \cite{Fujita:2007},\cite{Giessibl:2003}. However, common commercially available probes for AFM are made of silicon with a tip radius of app. 10nm, probes with tips of few nanometers in diameter are considered to be extremely sharp \cite{Giessibl:2003}. Silicon cantilevers possess a number of shortcomings, the fragility of tip apex and its degradation while scanning (probe ``wearing'') being the most noticeable ones.  To tackle these drawbacks carbon nanotube (CNT) probe was developed \cite{Dai:1996} by adhesion of a MWNT bundle to the tip of a silicon cantilever with a single nanotube protruding from the bundle last few hundred nanometers. Generally the diameter of MWNTs is in range of 5-30nm making MWNT probes not worse than silicon ones in terms of tip radius and having much higher aspect ratio. Moreover, it was shown \cite{Hafner:2001} that sub-nanometer resolution can be achieved by attaching a single-wall carbon nanotube (SWNT) to the tip apex. In general, CNT probes demonstrate following advantages over Si probes:
\begin{itemize}
\item CNT probes are more durable, preserve their shape while imaging \cite{Nguyen:2001};
\item if a cantilever with the CNT probe is metal-covered, it possesses sufficient conductivity to be used for Scanning Gate and Scanning Kelvin Probe Microscopy as well as to measure local voltage-current diagrams\cite{Chikashi:2002}. Metal-covered silicon cantilever shows substantial conductivity degradation during scanning due to destruction of the metal film around its tip apex while the CNT probe remains conductive during hours of continuous work;
\item if reliably attached, the CNT probe is not fragile compared to the silicon probe: under the excess load the CNT will bend while restoring its initial shape after the load is removed \cite{Dai:1996}.
\end{itemize}
Meanwhile, the CNT probe fabrication is a time-consuming task consisting of the following stages: a single CNT or a CNT bundle should be attached securely to the cantilever pyramid aligned with its axis or one of its faces. As the single CNT may be too long it should be shortened to achieve the desired rigidity. In case of the CNT bundle one should be sure that the bundle ends with a single nanotube to avoid a multiple tip. For the first time CNT probes were fabricated by coating the cantilever tip with an adhesive polymer and bringing it in contact with multiple MWNTs \cite{Dai:1996}. The authors succeeded to attach a SWNT bundle to the final MWNT, although they failed to achieve a single SWNT at its end. Later single SWNT as a probe was grown on the specifically prepared tip apex using Chemical Vapor Deposition \cite{Hafner:1999}. Finally, a simple and robust method of attaching a CNT bundle to the cantilever pyramid was developed exploiting dielectrophoresis phenomenon, pulling out the tip out of CNT suspension in water \cite{Tang:2005},\cite{Chikamoto:2013}.
It was shown \cite{Dai:1996},\cite{Hafner:2001gd} that by applying pulse voltage between the CNT probe and a conducting sample it is possible to shorten the final nanotube in a controllable way tens nanometers per pulse to achieve desired probe rigidity. In \cite{Ito:2003} the process of electro-chemical etching was used to shorten MWNTs of big radius (app. 200nm) in an aqueous electrolyte solution applying DC of app. 2V. It was shown that the part of the nanotube immersed into the electrolyte was etched completely while the rest of the nanotube above the electrolyte surface was remained intact. This method can be used to shorten a nanotube or a nanotube bundle on scales of 1$\mu m$ and more, controlling the process in an optical microscope (a single MWNT or a CNT bundle of 200nm in diameter are visible in an optical microscope).
Knowledge of the probe shape may provide useful information while analyzing an SPM image. The most reliable instrument for that is Scanning Electron Microscope or Transmitting Electron Microscope (SEM/TEM). However, frequently it's more convenient to obtain this information directly in SPM taking into account that probe shape may change while imaging. Scanning a known-beforehand test-structure provides such probe characteristics as its apex radius and aspect ratio. It's shown \cite{Fujita:2007} that a rigid sharp protrusion on a flat surface is an optimal test-structure to restore the tip surface function. However, fabrication of the protrusion with radius at its tip smaller or comparable to the sharp SPM probe is a complex technological task \cite{Fujita:2007}. In practice, common test-structures for probe shape characterization are nano-objects deposited on a flat surface like gold sol nanoparticles or CNTs as well as a surface transformed to a regular structure like an array of pyramids or comb-like structures with peak radius of about 10nm. However, fabrication of such structures requires complex technological methods. In \cite{Sun:2007},\cite{Sui:2001} AFM methods are used to study alumina obtained by anodization. It's shown that carefully selecting anodization parameters a set of alumina peaks with radius at apexes of about 10nm can be obtained. As the interface between the alumina and aluminium has a developed structure, the surface of aluminium after selective alumina removal (the dimpled aluminium) is shown to be a suitable test-structure for tip shape characterization \cite{Belov:2010}. The radius of peaks on the surface was estimated to be 1-2nm.
Considering the test-structures available for probe shape characterization we chose the dimpled aluminium as the object close to optimal one and which can be relatively easy fabricated. Under some anodization parameters the surface of the dimpled aluminium is a set of dimples ordered in a hexagonal structure, depicted schematically in Figure \ref{fig:fig1a}. At the intersection points of three neighboring walls of the dimples peaks with radius less than 10nm are formed. Such highly ordered structure can be observed after selective removal of alumina \cite{Zhang:1998} as shown in Figure \ref{fig:fig1b}.
Typical lattice constant is 100-200nm, the height difference between the peak apex and the dimple bottom is 20-50nm (the lattice constant and peaks height depend on the anodization parameters). Scanning such an object in AFM reveals probe shape information about its last 20-50nm.
Herein we describe a relatively simple procedure of the MWNT probe fabrication with a single nanotube at its end on the pyramid of a silicon cantilever applying dielectrophoresis technique to the MWNT suspension in water. To transform the CNT bundle attached to the pyramid into a probe suitable for SPM imaging we use the dimpled aluminium sample as a test-structure for tip shape determination as well as for the CNT bundle shortening/modification applying pulse voltage between the bundle and the sample. The probe produced in this way ends with a single nanotube reliably attached to the cantilever capable of AFM imaging with resolution not worse than that of the silicon cantilever. If Si cantilevers are metal-covered, the majority of the CNT probes produced on them turn out to be conductive.

\section{Experimental methods}
\begin{figure}[ht]\centering
\includegraphics[width=\linewidth]{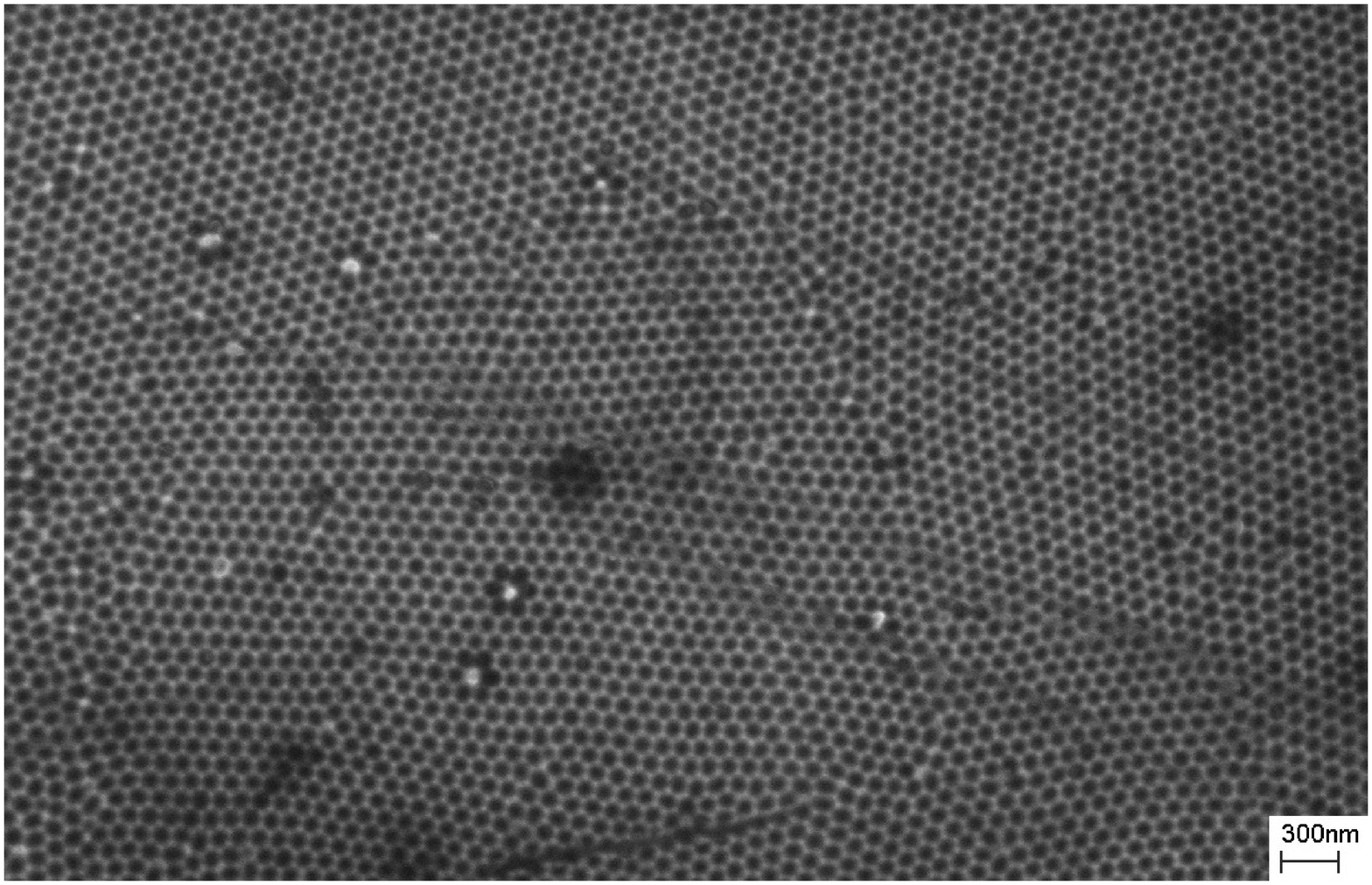}
\caption{SEM image of the dimpled aluminium reveals long-range order}
\label{fig:fig2}
\end{figure}

\begin{figure}[ht]
        \centering
        \begin{subfigure}[b]{0.45\linewidth}
                \includegraphics[width=\linewidth]{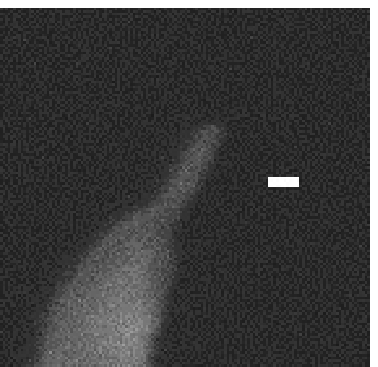}
                \caption{}
                \label{fig:fig3a}
        \end{subfigure}%
        ~
        \begin{subfigure}[b]{0.447\linewidth}
                \includegraphics[width=\linewidth]{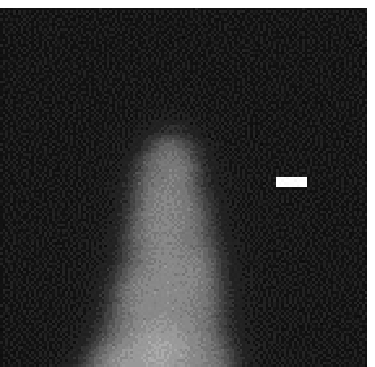}
                \caption{}
                \label{fig:fig3b}
        \end{subfigure}
        ~
        \begin{subfigure}[b]{0.45\linewidth}
                \includegraphics[width=\linewidth]{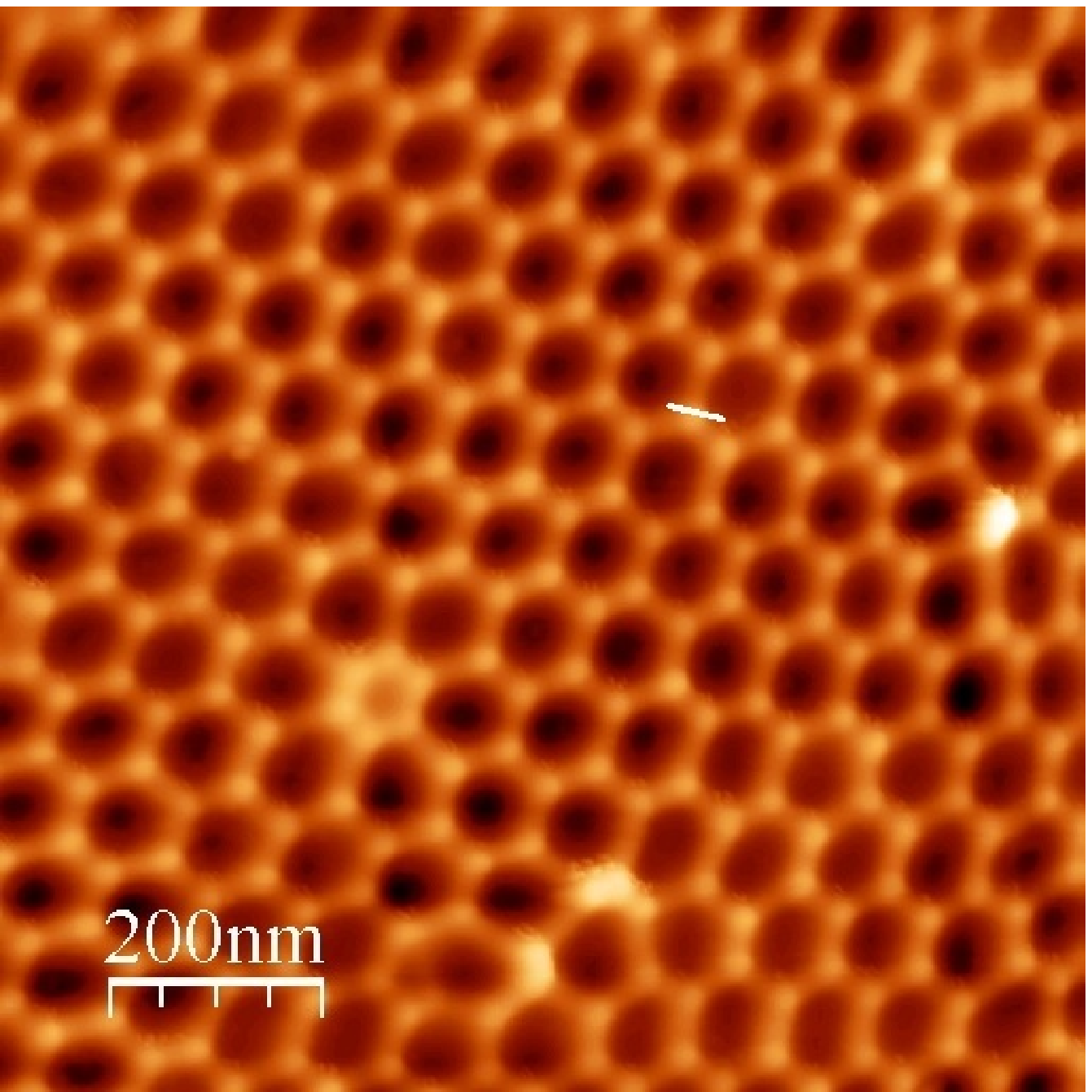}
                \caption{}
                \label{fig:fig3c}
        \end{subfigure}%
        ~
        \begin{subfigure}[b]{0.45\linewidth}
                \includegraphics[width=\linewidth]{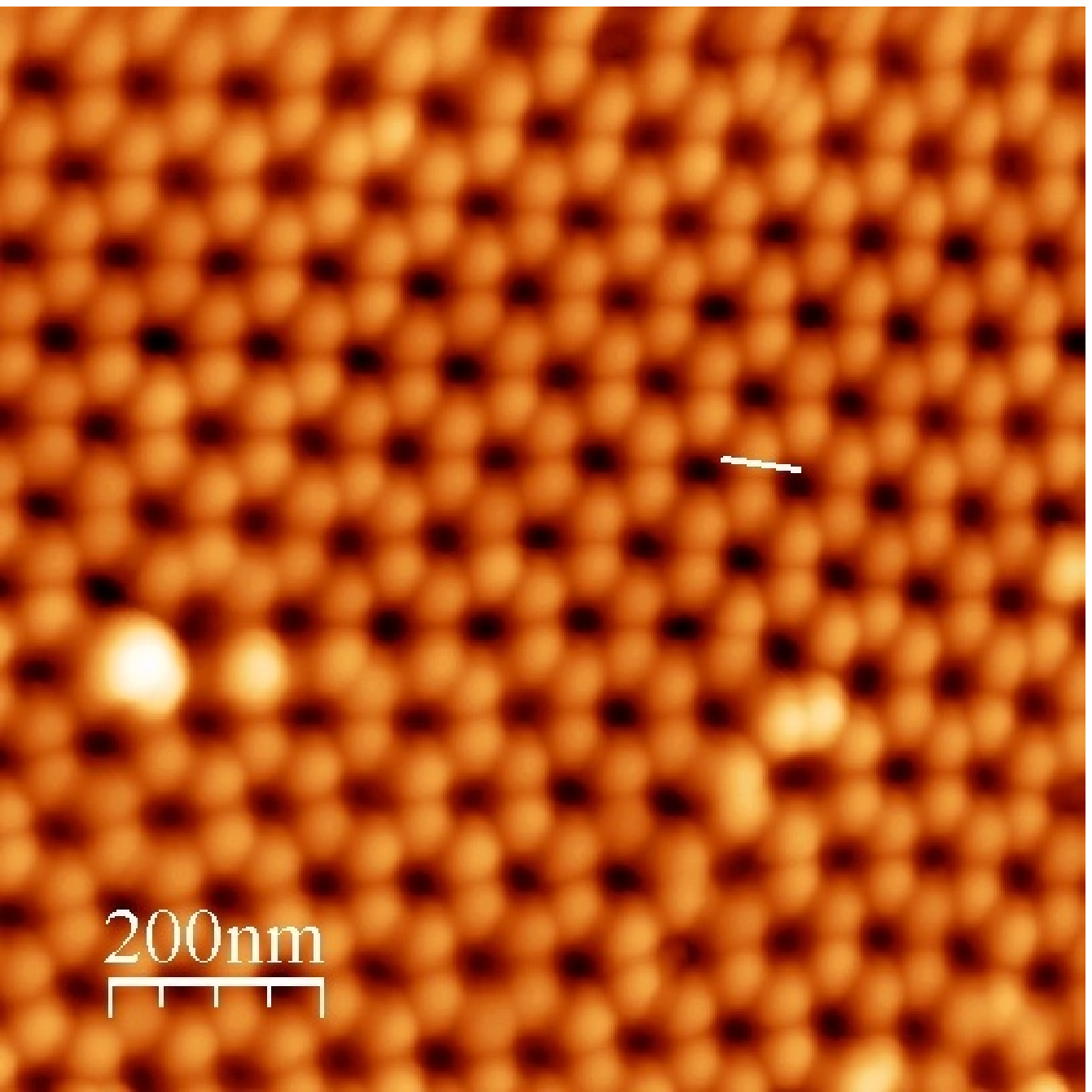}
                \caption{}
                \label{fig:fig3d}
        \end{subfigure}
        \begin{subfigure}[b]{\linewidth}
                \includegraphics[width=\linewidth]{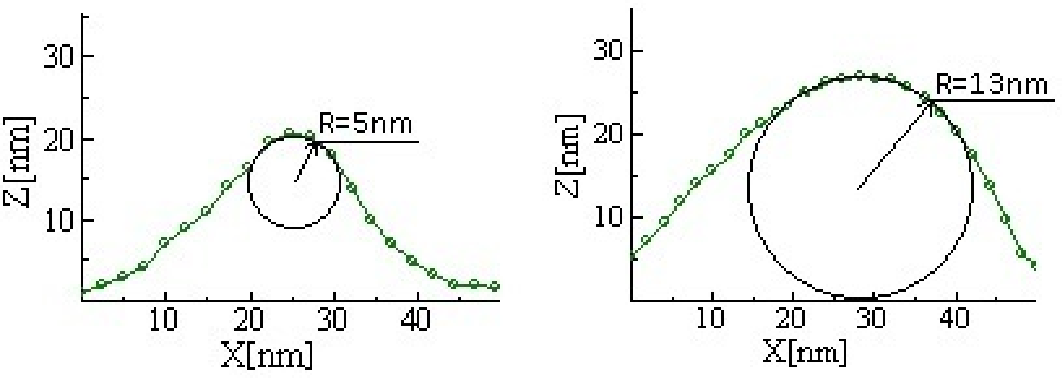}
                \caption{}
                \label{fig:fig3e}
        \end{subfigure}%
        \caption{SEM images of (a) a diamond probe and (b) a Si probe, the scale bar is 20nm. AFM images of the dimpled aluminium obtained by (c) the diamond probe and (d) the Si probe. The height of the peaks is 25nm in both images.  (e) Estimates of the probes radius obtained by building profiles using the AFM images.}
        \label{fig:fig3}
\end{figure}

In our work the dimpled aluminium sample was provided by MSU chemical department. It's a round piece of aluminium foil with a diameter of 9mm, 6mm of which are covered with dimples ordered hexagonally as described above. A typical region of 5x3$\mu m$ is shown in Figure \ref{fig:fig2} obtained by SEM. Though there are clearly defects on the long range, the structure is nearly perfectly ordered hexagonally. The sample is homogenous almost on the entire surface covered with dimples.
The lattice constant of the structure is $105\pm5$nm. AFM imaging of the sample shows that the dimples depth is in range of 25-35nm.
The main advantage of this test-structure is that it allows to perform express evaluation of an AFM probe shape including radius estimation without necessity to directly investigate it in SEM/TEM. To illustrate the idea two cantilevers where chosen:
\begin{enumerate}
\item a silicon cantilever, tip radius $13\pm3$nm (MikroScience);
\item a cantilever with the diamond pyramid with tip radius of $5\pm2$nm (ART\texttrademark Single Crystal Diamond).
\end{enumerate}
To compare these probes they were studied in SEM as well as they were used to scan the dimpled aluminium sample in AFM. The imaging was made on the atomic force microscope NTEGRA (NT-MDT, Russia) on air, the area of scanning 1x1$\mu m$ with scan speed of 1Hz. Different cross-sections of the sample surface were built. Profiles with the least radius of curvature where found to have the radius of 5nm and 13nm for (1) and (2) cantilevers respectively. These values provide an estimate of the sum of the probe and surface profile radii. In \cite{Belov:2010} similar analysis gave the estimate of the radius of the dimples walls cross-sections of 1-2nm for their sample of the dimpled aluminium. Taking into account of our SEM and AFM images we arrived at the same estimate for the peaks of our sample. Based on this result it was concluded that the dimpled aluminium is an effective instrument for an AFM probe shape characterization if the probe radius is 2nm and more. The results are shown in Figure \ref{fig:fig3}.
To get qualitative information on the tip sharpness, visual shape and height of the dimple aluminium peaks in the AFM image should be considered. For the high aspect ratio probe with tip radius of less than 10nm the visual peaks diameter was an order of magnitude smaller compared to the lattice constant of our sample and the peaks height was 20-30nm as shown in Figure \ref{fig:fig3c}. The multiple tip apex can be immediately diagnosed based on the multiple peaks shape like in Figure \ref{fig:fig5a}.

\section{CNT probes fabrication}
\begin{figure}[ht]
        \centering
        \begin{subfigure}[b]{0.48\linewidth}
                \includegraphics[width=\linewidth]{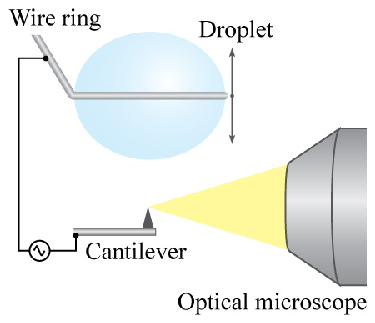}
                \caption{}
                \label{fig:fig4a}
        \end{subfigure}%
        ~
        \begin{subfigure}[b]{0.48\linewidth}
                \includegraphics[width=\linewidth]{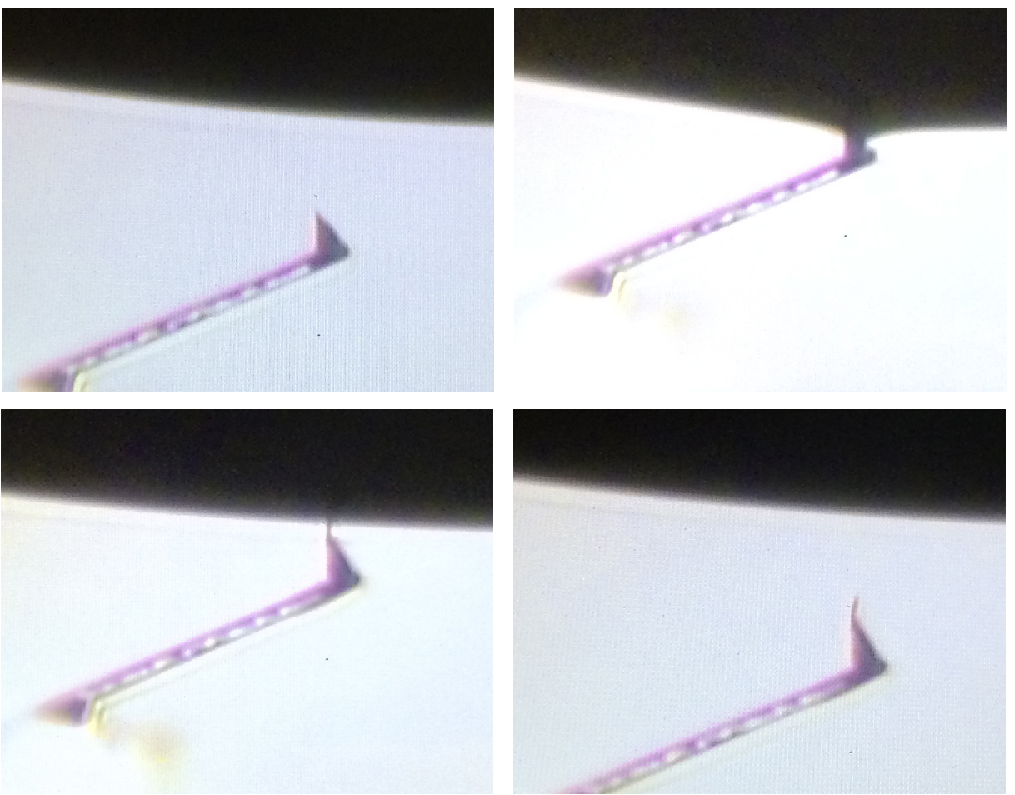}
                \caption{}
                \label{fig:fig4b}
        \end{subfigure}
        \caption{(a) Scheme of a setup for a MWNT bundle formation on the tip of a cantilever. (b) Acquired optical image of the four stages of the bundle formation.}
        \label{fig:fig4}
\end{figure}

\begin{figure}[ht]
        \centering
        \begin{subfigure}[]{0.439\linewidth}
        \centering
                \includegraphics[width=\linewidth]{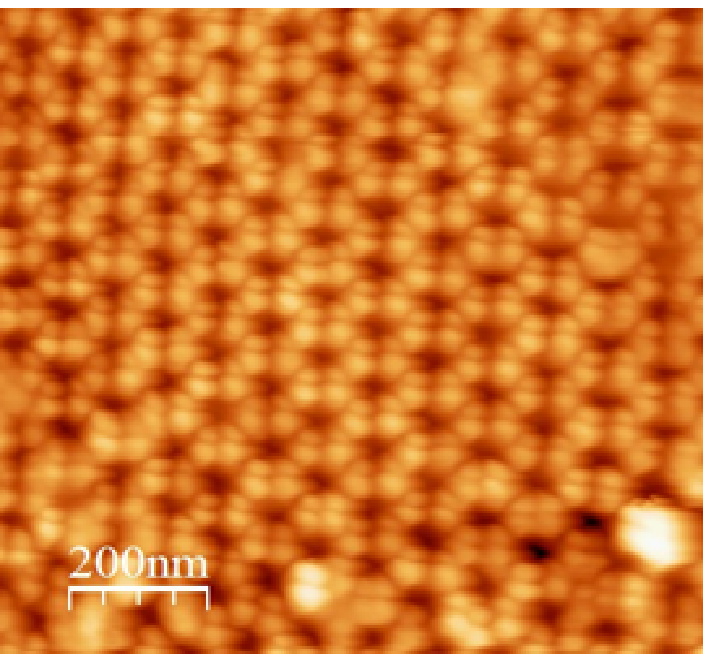}
                \caption{}
                \label{fig:fig5a}
        \end{subfigure}%
        ~
        \begin{subfigure}[]{0.4\linewidth}
        \centering
                \includegraphics[width=\linewidth]{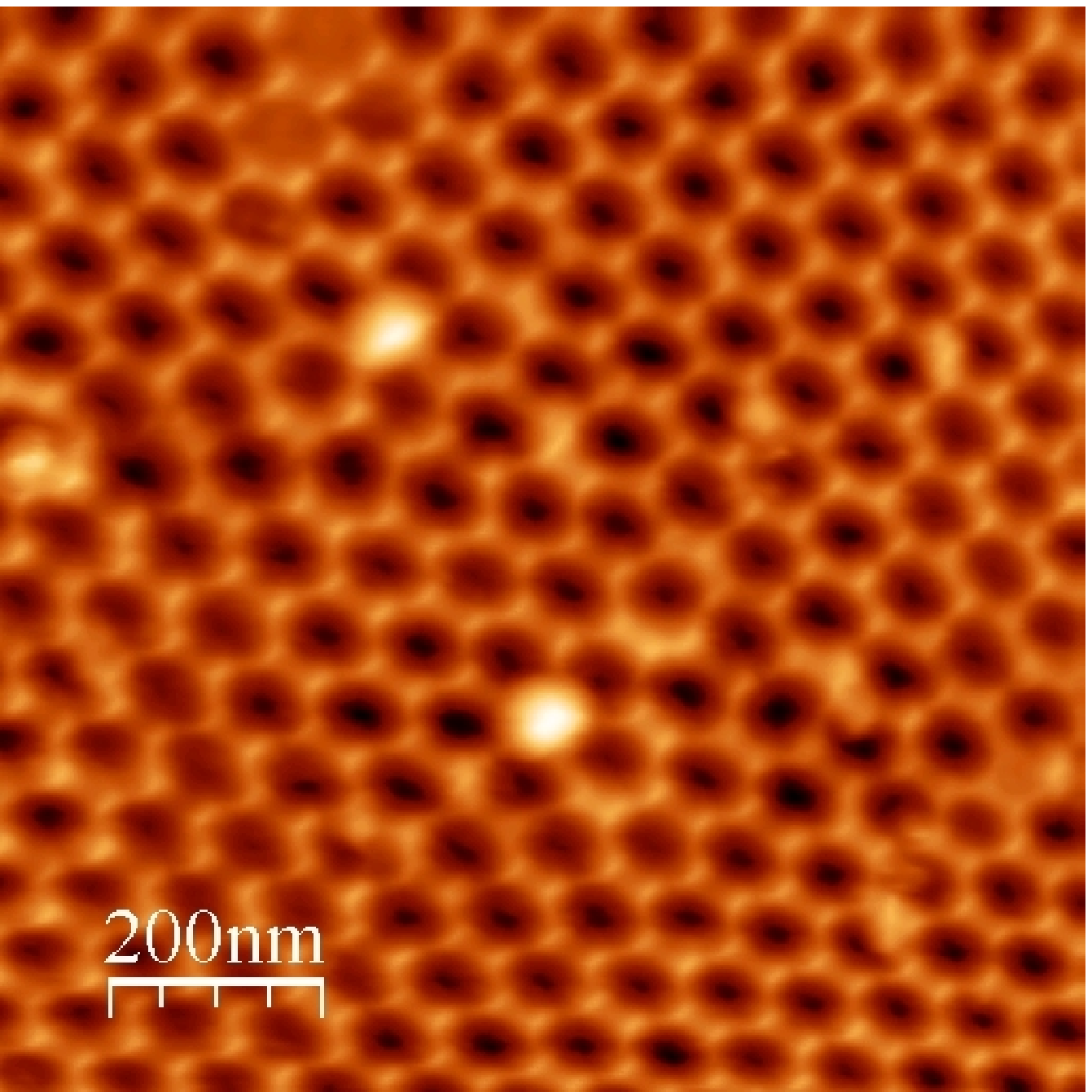}
                \caption{}
                \label{fig:fig5b}
        \end{subfigure}
        \caption{(a) AFM scan of the dimpled aluminium revealing multiple probe apex. The height of the peaks is 15nm. (b) Etalon image of the dimpled aluminium by a MWNT probe. The height of the peaks is 30nm.}
        \label{fig:fig5}
\end{figure}

To form a CNT bundle on a cantilever pyramid the method of dielectrophoresis deposition was chosen. The scheme of our setup is shown in Figure \ref{fig:fig4a}.
To prepare a MWNT suspension 50mg of MWNTs (Sigma-Aldrich, CAS Number 308068-56-6, product number 406074, 1-10$\mu m$ long, 15-20nm in diameter) and 5ml of distilled water was used. A droplet of the suspension was placed on a platinum wire ring (ring diameter 2mm, wire diameter 0.25mm). An AC voltage of 5-10V, 2MHz was applied between the wire and the cantilever. Under an optical microscope observation the tip of the cantilever pyramid was immersed into the suspension using a fine-adjustment screw 2-5$\mu m$ deep and slowly pulled out. The dielectrophoresis force drags CNTs towards the tip, forming a CNT bundle 1-10$\mu m$ long reliably attached to the tip (Figure \ref{fig:fig4b}). Forces of surface tension in the meniscus of suspension around the tip align nanotubes into the bundle. The angle between the pyramid axis and the CNT bundle is defined by joint orientation of the pyramid and surface of the suspension and has distribution of less than 12$^{\circ}$ \cite{Tang:2005}. In some cases the bundle was not formed or appeared to be not visible in our microscope during the first try. Repeating the procedure one can succeed in forming the CNT bundle. Investigation of bundles obtained this way in SEM revealed that a bundle of 200nm in diameter was already visible in our optical microscope. This procedure of attaching nanotubes is applicable to MWNTs as well as to SWNTs \cite{Tang:2005}.
AFM imaging using CNT probes showed that in most cases the probes possessed longitudinal or transverse instability, if the bundle was 10$\mu m$ long or more. For this reason just after the bundle formation its length was estimated in the optical microscope. If the visual length was more than 3$\mu m$ (as was in most cases), the bundle was shortened after being annealed on air at 120$^{\circ}$C for 30min. For shortening electro-chemical etching was used in the same setup where we had the second wire ring to keep a droplet of KOH. Applying 4V, 0.01-1msec pulses it's possible to shorten the bundle length in steps of 1$\mu m$ and more by etching immersed part of the bundle. SEM imaging of etched bundles showed that the bundle extended 2-5$\mu m$ from the pyramid and in most cases ended with a single nanotube with length of up to few micrometers. 
The next step is to check AFM imaging capabilities of the CNT probe. If the image of the dimpled aluminium obtained by the probe looks as it's supposed to be (Figure \ref{fig:fig5b}), and the probe is sufficiently rigid, which is checked using force curves, the probe is ready for work. Otherwise, the probe is not sufficiently rigid or reveals a multiple apex similar to Figure \ref{fig:fig5a}.
In this case the probe should be shortened/modified. To do it in a more controllable way than electrochemical etching, pulse voltage (30-50V, 10-100$\mu sec$) was applied between the cantilever and the dimpled aluminium sample. The current was limited by a resistor of 1MOhm inserted in series. Using this method the probe length was shortened by 50-500nm per pulse depending on the pulse parameters with feedback turned on in the semi-contact mode. Typically few pulses were enough to gain required rigidity and single apex of the MWNT probe. In Figure 6 typical MWNT probes ready for AFM imaging are shown.
\begin{figure}[tbp]
        \centering
        \begin{subfigure}[]{0.35\linewidth}
        \centering
                \includegraphics[width=\linewidth]{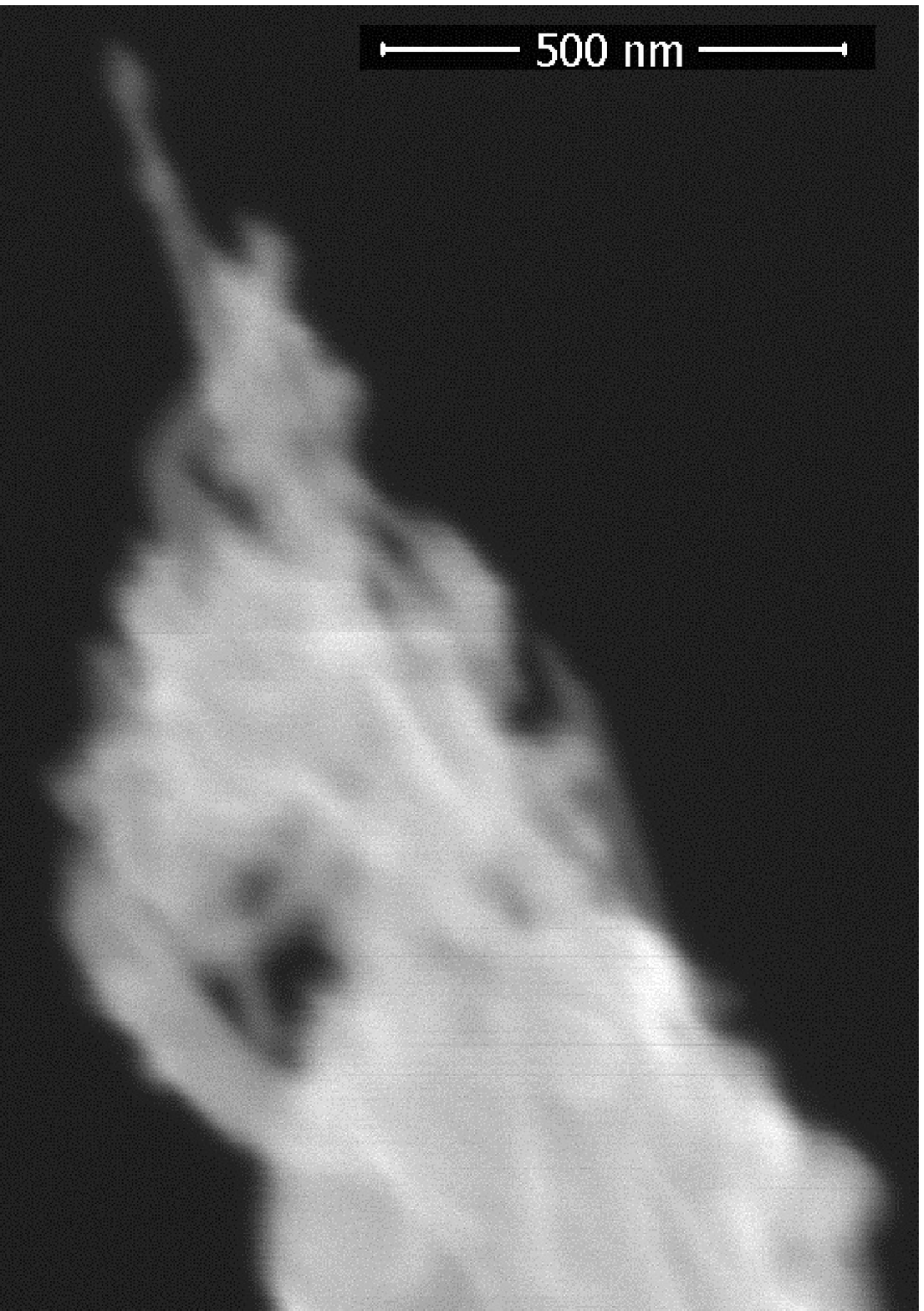}
                \caption{}
                \label{fig:fig6a}
        \end{subfigure}%
        ~
        \begin{subfigure}[]{0.322\linewidth}
        \centering
                \includegraphics[width=\linewidth]{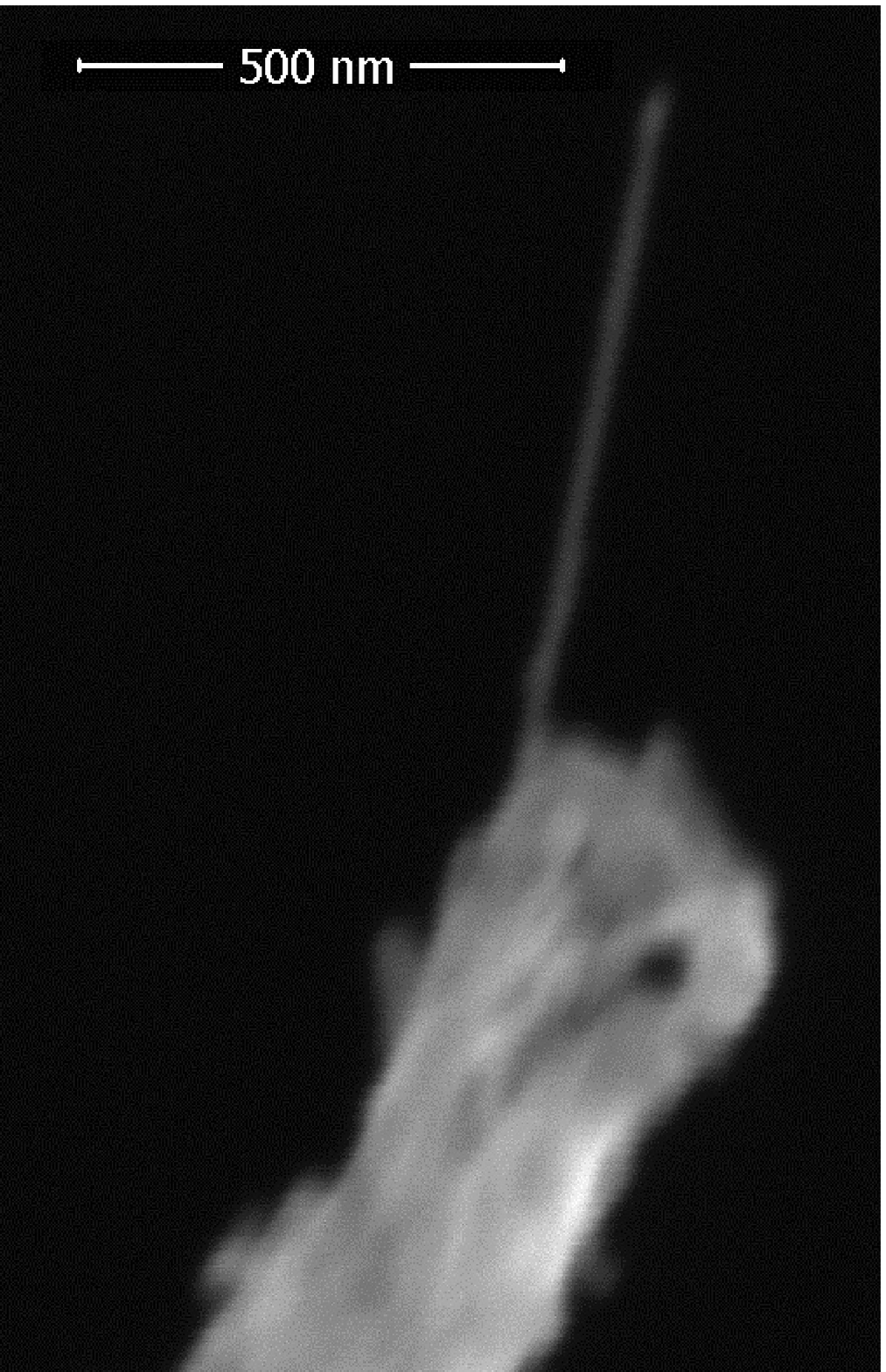}
                \caption{}
                \label{fig:fig6b}
        \end{subfigure}
        \caption{ MWNT probes ready for SPM imaging. A single nanotube protrudes from the bundle. The bundles contain lots of amorphous Carbon.}
        \label{fig:fig6}
\end{figure}
Ready MWNT probes contained high amount of amorphous carbon and many nanotubes were not aligned to the probe axis, which can be seen from the SEM images (Figure \ref{fig:fig6}). To obtain CNT probes without amorphous carbon the MWNTs should be purified prior to the dielectrophoresis step. However, the presence of amorphous Carbon in the probe doesn't affect quality of AFM images. Study of 20 MWNT probes in SEM revealed that the length of the nanotube part protruded from the bundle is more than 40nm in half of the cases.
Initial studies of the MWNT probes produced by this method revealed that the majority of probes were conductive. The local conductivity of a deposited gold-palladium film was studied using one of the probes. The probe was brought in contact with the sample to the pressure of 5nN on air and voltage-current curve was measured at each point of scanning 1x1$\mu m$. Topography, map of current at probe potential of +0.3V and the averaged V-I curve for the outlined area are shown in Figure \ref{fig:fig7}.
\begin{figure}[ht]
        \centering
        \begin{subfigure}[b]{0.45\linewidth}
                \includegraphics[width=\linewidth]{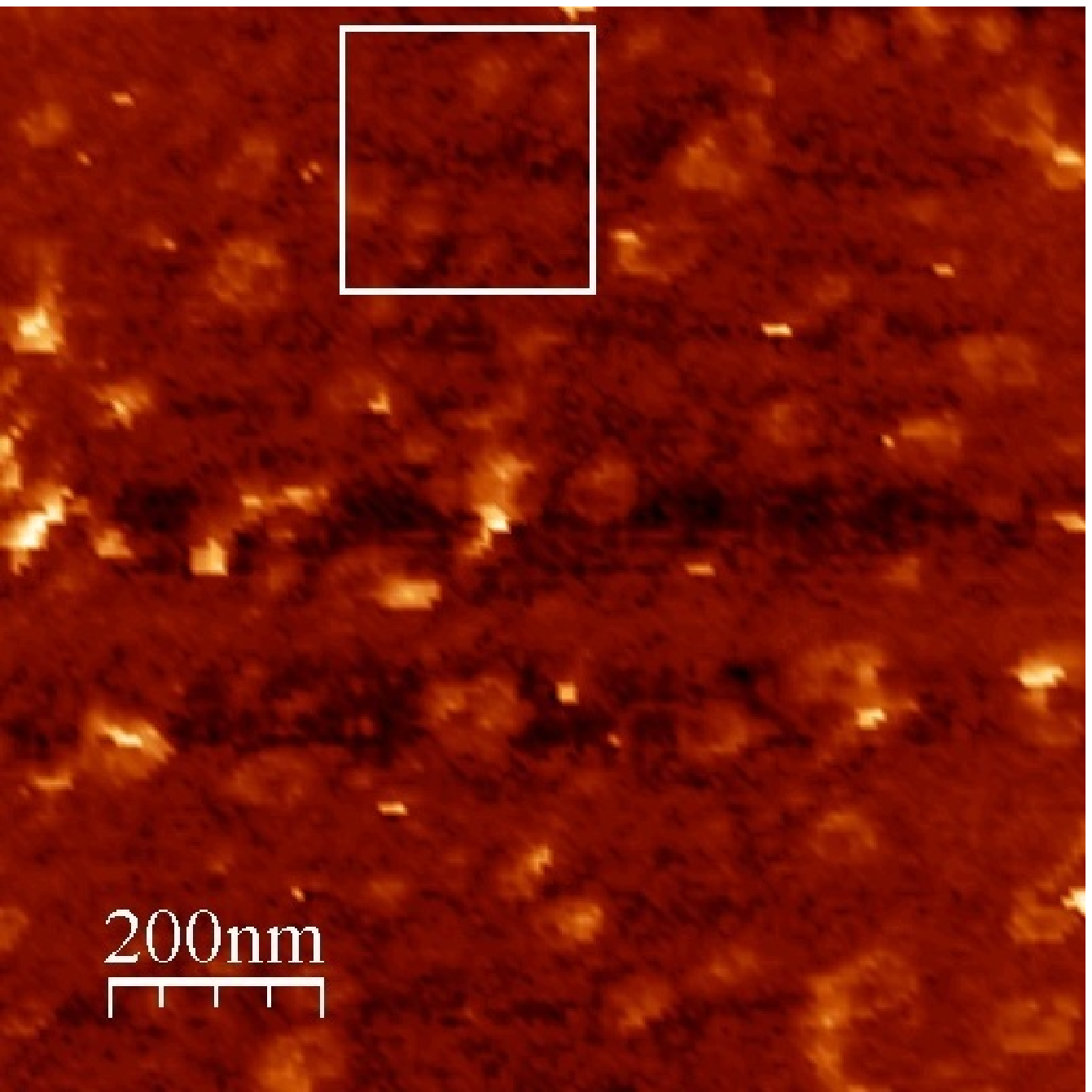}
                \caption{}
                \label{fig:fig7a}
        \end{subfigure}%
        ~
        \begin{subfigure}[b]{0.45\linewidth}
                \includegraphics[width=\linewidth]{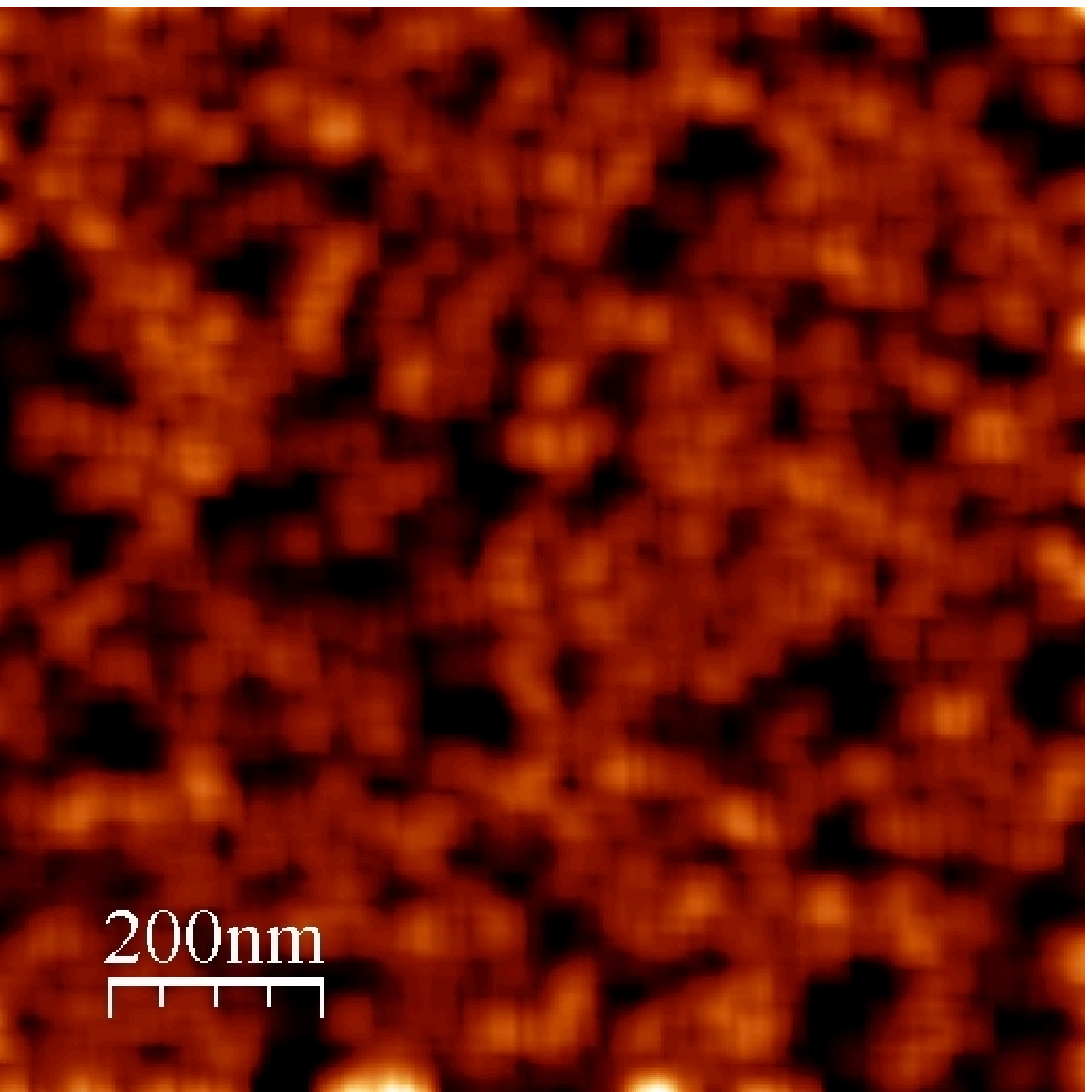}
                \caption{}
                \label{fig:fig7b}
        \end{subfigure}

        \begin{subfigure}[b]{0.8\linewidth}
                \includegraphics[width=\linewidth]{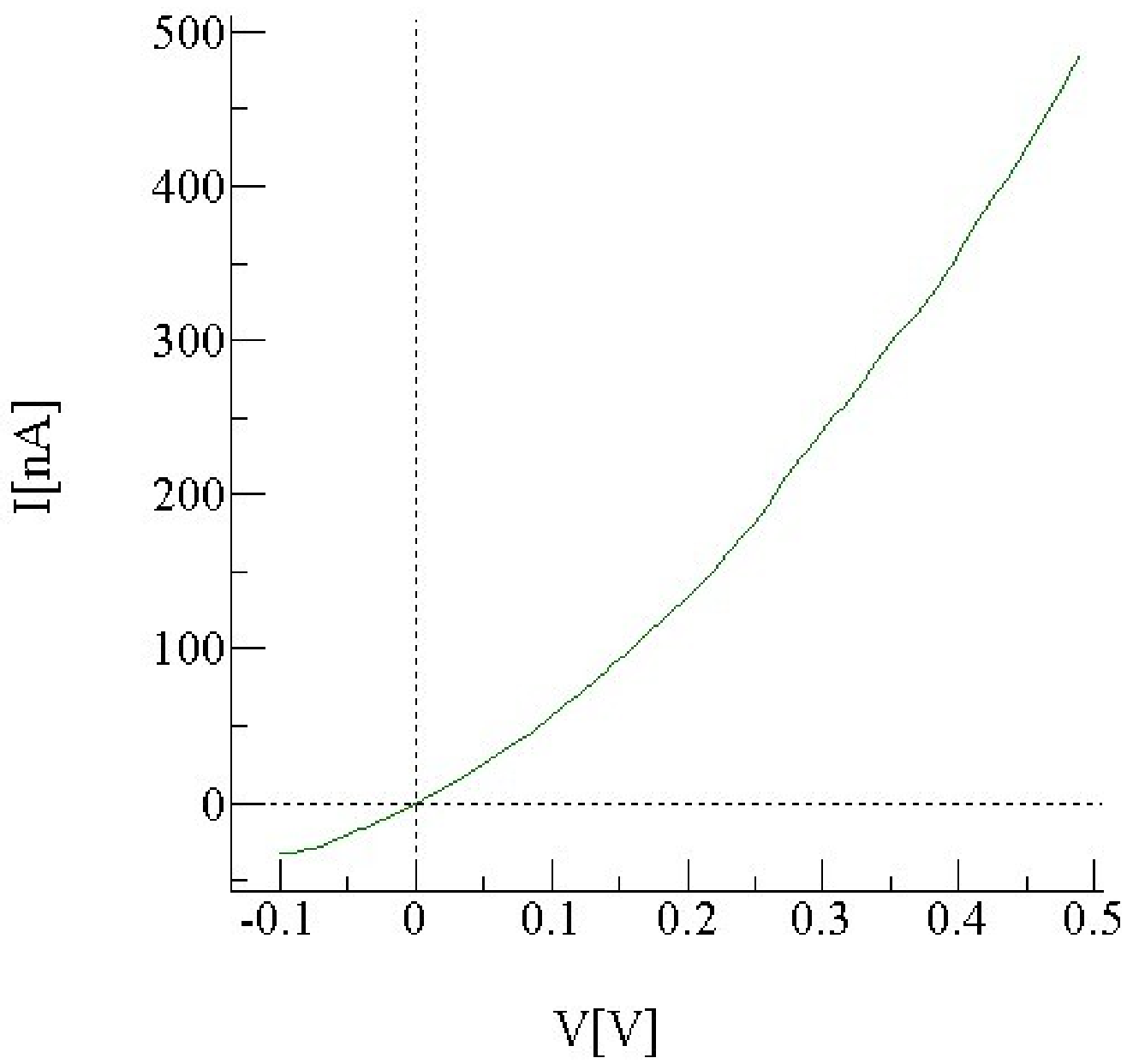}
                \caption{}
                \label{fig:fig7b}
        \end{subfigure}
        \caption{(a) Topography of a gold-palladium film obtained by a MWNT probe in a modified contact mode, the height range is 0-40nm. (b) The smoothed current map obtained during the scan at constant force of 5nN and the probe potential of +0.3V. The current range is -200 .. 1800nA. (c) The averaged V-I curve from the outlined area from (a).}
        \label{fig:fig7}
\end{figure}
Repeated scanning of the sample during 2 hours revealed no change in the current map indicating no conductivity degradation of the MWNT probe.
\section{Conclusions}
Here in we demonstrated the method of MWNT probe fabrication on the pyramid of Si cantilevers. To form a nanotube bundle on the tip of the pyramid the electrophoresis technique was used. Express method of an AFM tip shape characterization was shown via imaging the dimpled aluminium sample in AFM. This test structure was used to transform the CNT bundle into an SPM probe capable of imaging in the contact and modulation modes. Electrochemical etching in an aqueous KOH solution was utilized for the initial bundle shortening. Pulse voltage was applied between the sample of the dimpled aluminium and the bundle in the tapping mode to form a single-nanotube probe rigid enough to work in the contact mode. The pulse either shortens the last too long nanotube or burns few last nanotubes of the same length in case of multiple apex together with the amorphous carbon. AFM scanning of the same sample was used to control changes occurred to the CNT probe.
The majority of the MWNT probes fabricated on the cantilevers covered with gold were found to be conductive. It was demonstrated studying the local conductivity of a gold-palladium film sputtered on a Si substrate. The shape and conductivity of the produced probes degraded much slower compared to gold-covered Si cantilevers.


\section{Acknowledgments }
We are grateful to P.A. Malyshkin (Artech Carbon OU) for supplying us with cantilevers with the diamond pyramid and K.S. Napolsky (Moscow State University) for providing samples of the dimpled aluminium. This work was supported by megagrant No.14Y26.31.0007 funded by Russian Ministry of Education and Science, and in part by RAS, RFBR grant No. 12-02-00573.


\end{document}